\documentclass[11pt]{article}
\usepackage{latexsym}
\usepackage{amssymb}

\newtheorem{lemma}{Lemma}

\def\qed{\ \hfill$\Box$}

\title{{\bf Coloring vertices of a graph or finding a Meyniel
obstruction}\thanks{This work was partially supported by the Algorithmic
Discrete Optimization Network (ADONET), the Natural Sciences and
Engineering Research Council of Canada (NSERC), and the Research
Grants Program of Wilfrid Laurier University}}

\author{Kathie Cameron\thanks{Laurier University, Waterloo, Ontario,
    Canada N2L 3C5, kcameron@wlu.ca}, Jack Edmonds\thanks{EP Institute,
    Kitchener, Ontario, Canada, N2M 2M6, jackedmonds@rogers.com},
  Benjamin L\'ev\^eque\thanks{Laboratoire G-SCOP, 46 avenue F\'elix
    Viallet, 38031 Grenoble Cedex, France,
    benjamin.leveque@g-scop.inpg.fr}, Fr\'ed\'eric
  Maffray\thanks{C.N.R.S., Laboratoire G-SCOP, 46 avenue F\'elix
    Viallet, 38031 Grenoble Cedex, France, frederic.maffray@g-scop.inpg.fr}}

\begin{document}

\maketitle
\begin{abstract}
A Meyniel obstruction is an odd cycle with at least five vertices and
at most one chord.  A graph is Meyniel if and only if it has no
Meyniel obstruction as an induced subgraph.  Here we give a $\mathcal
O(n^2)$ algorithm that, for any graph, finds either a clique and
coloring of the same size or a Meyniel obstruction.  We also give a
$\mathcal O(n^3)$ algorithm that, for any graph, finds either an
easily recognizable strong stable set or a Meyniel obstruction.

\end{abstract}

\section{Introduction}

A graph is \emph{Meyniel} \cite{mey0} if every odd cycle of length at
least five has at least two chords.  A \emph{Meyniel obstruction} is
an odd cycle of length at least five with at most one chord.  Thus a
graph is Meyniel if and only if it does not contain a Meyniel
obstruction as an induced subgraph.  Meyniel \cite{mey0} and Markosyan
and Karapetyan \cite{MarKar} proved that Meyniel graphs are perfect.
This theorem can be stated in the following way:
\begin{quote}
{\it For any graph $G$, either $G$ contains a Meyniel obstruction, or
$G$ contains a clique and coloring of the same size (or both).}
\end{quote}

We give a polytime algorithm which finds, in any graph, an instance of
what the Meyniel-Markosyan-Karapetyan theorem says exists.  This
algorithm works in time $\mathcal O(n^2)$ where $n$ is the number of
vertices of the input graph.  This is an improvement in the complexity
of the algorithm of the first and second authors
\cite{camedmSIAM,camedmMeyniel}, which finds, in any graph, a clique
and coloring of the same size, or a Meyniel obstrution.  This is an
enhancement of the $\mathcal O(n^2)$ algorithm of Roussel and Rusu
\cite{rourus01}, which optimally colors any Meyniel graph.

This work is motivated by the ``Perfect Graph Robust Algorithm
Problem'' \cite{camedm}: seek a polytime algorithm which, for any
graph $G$, finds either a clique and a coloring of the same size or an
easily recognizable combinatorial obstruction to $G$ being perfect.
According to the Strong Perfect Graph Theorem \cite{CRST}, proved in a
different way, a simple obstruction to perfectness is the existence of
an odd hole or odd antihole.

A \emph{stable set} in a graph $G$ is a set of vertices, no two of
which are joined by an edge of $G$.  A \emph{strong stable set} in $G$
is a stable set that contains a vertex of every maximal (by inclusion)
clique of $G$.  Note that if one can find a strong stable set in every
induced subgraph of a graph $G$, one can easily find an optimal
coloring of $G$ : if $S_1$ is a strong stable set of $G$, $S_2$ is a
strong stable set of $G\backslash S_1$, $\ldots$, $S_\ell$ is a strong
stable set of $G\backslash (S_1 \cup \ldots \cup S_{\ell-1})$, and $S_\ell$
is the last non-empty such set, then $S_1,\ldots,S_{\ell}$ is a coloring
of $G$ which is the same size as some clique of $G$.

Ravindra \cite{rav} presented the theorem that
\begin{quote}
{\it For any graph $G$, either $G$ contains a Meyniel obstruction, or
$G$ contains a strong stable set (or both).}
\end{quote}

Ravindra's proof is an informal description of an algorithm which
finds, in any graph, an instance of what the theorem says exists.

Ho\`ang \cite{hoa} strengthened this to the following:
\begin{quote}
{\it For any graph $G$ and vertex $v$ of $G$, either $G$ contains a
Meyniel obstruction, or $G$ contains a strong stable set containing
$v$ (or both).}
\end{quote}

Hoang \cite{hoa} give a $\mathcal O(n^7)$ algorithm that finds, for
any vertex of a Meyniel graph, a strong stable set containing this
vertex.

A disadvantage of the Ravindra-Ho\`ang theorem is that it is not an
existentially polytime theorem.  A theorem is called
\emph{existentially polytime (EP)} if it is a disjunction of NP
predicates which is always true \cite{camedm}.  The predicate ``$G$
contains a strong stable set" may not be an NP-predicate because the
definition of strong stable set is not a polytime certificate.

The Ravindra-Ho\`ang theorem is strengthened in
\cite{camedmSIAM,camedmMeyniel} to:
\begin{quote}
{\it For any graph $G$ and vertex $v$ of $G$, either $G$ contains a
Meyniel obstruction, or $G$ contains a nice stable set containing
$v$ (or both),}
\end{quote}
where nice stable sets are a particular type of strong stable set
which have the following polytime-certifiable meaning.  A \emph{nice
stable set} in a graph $G$ is a maximal stable set $S$ linearly
ordered so that there is no induced $P_4$ between any vertex $x$ of
$S$ and the vertex which arises from the contraction in $G$ of all
the vertices of $S$ that precede $x$. (\emph{Contracting} vertices
$x_1, \ldots, x_k$ in a graph means removing them and adding a new
vertex $x$ with an edge between $x$ and every vertex of
$G\setminus\{x_1, \ldots, x_k\}$ that is adjacent to at least one of
$x_1, \ldots, x_k$.  As usual, $P_4$ denotes a path on four
vertices.) The proof of the theorem in
\cite{camedmSIAM,camedmMeyniel} is a polytime algorithm which for
any graph and any vertex in that graph, finds an instance of what
the theorem says exists.  In Section \ref{stable}, we give an
${\mathcal O}(n^3)$ algorithm for this, where $n$ is the number of
vertices of the input graph.

%%%%%%
\section{The coloring algorithm}

We recall the algorithm {\sc LexColor} of Roussel and Rusu
\cite{rourus01} which is a ${\mathcal O}(n^2)$ algorithm that colors
optimally the vertices of a Meyniel graph, thereby improving the
complexity of previous coloring algorithms due to Hertz ${\mathcal
O}(nm)$ \cite{her} (where $m$ is the number of edges of the input
graph), Ho\`ang ${\mathcal O}(n^8)$ \cite{hoa} and Ravindra
\cite{rav}.

{\sc LexColor} is a greedy coloring algorithm.  The integers $1, 2,
\ldots, n$ are viewed as colors.  For each vertex $x$ of $G$ and each
color $c\in \{1, 2, \ldots, n\}$, we have a label $label_x(c)$ defined
as follows.  If $x$ has no neighbor colored $c$, then $label_x(c)$ is
equal to $0$; if $x$ has a neighbor colored $c$, then $label_x(c)$ is
equal to the integer $i$ such that the first neighbor of $x$ colored
$c$ is the ($n$-$i$)-th colored vertex of the graph.  We consider the
following (reverse) lexicographic order on the labels : $label_x
<_{Lex} label_y$ if and only if there exists a color $c$ such that
$label_x(c)<label_y(c)$ and $\forall\ c' > c$,
$label_x(c')=label_y(c')$.
At each step, the algorithm selects an uncolored vertex which is
maximum for the lexicographic order of the labels, assigns to this
vertex the smallest color not present in its neighbourhood, and
iterates this procedure until every vertex is colored.  More formally:

\begin{quote}
{\sc Algorithm LexColor}

{\it Input:} A graph $G$ with $n$ vertices.

{\it Output:} A coloring of the vertices of $G$.

%\textit{Method:}

{\it Initialization:}
For every vertex $x$ of $G$ and every color $c$ do $label_x(c) := 0$;

{\it General step:} For $i=1, \ldots, n$ do:\\
1.  Choose an uncolored vertex $x$ that maximizes $label_x$ for
$<_{Lex}$;\\
2.  Color $x$ with the smallest color $c$ not present in its
neighbourhood;\\
3.  For every uncolored neighbor $y$ of $x$, if $label_y(c) := 0$ do
$label_y(c) := n-i.$
\end{quote}

This coloring algorithm is optimal on Meyniel graph and its complexity
is ${\mathcal O}(n^2)$ \cite{rourus01}.

\

\emph{Remark 1:} This version of {\sc LexColor} has a minor
modification from the original version of Roussel and Rusu
\cite{rourus01} .  When $x$ has a neighbor colored $c$, the integer
$label_x(c)$ was originally defined to be the integer $i$ such that
the first neighbor of $x$ colored $c$ is the ($n-i$)-th vertex colored
$c$ of the graph (instead of the ($n-i$)-th colored vertex of the
graph).  For a color $c$, the order between $label_x(c)$ of each
vertex $x$ is the same in the two versions of the algorithm, so the
lexicographic order is the same and there is no difference in the two
executions of the algorithm.  This modification only simplifies the
description of the algorithm.

\

\emph{Remark 2:} The graph $\overline{P}_6$ is an example of a
non-Meyniel graph on which Algorithm {\sc LexColor} is not optimal.
The graph $\overline{P}_6$ has vertices $u, v, w, x, y, z$ and
non-edges are $uv, vw, wx, xy, yz$.  Algorithm {\sc LexColor} can
color the vertices in the following order with the indicated color:
$v-1$, $y-2$, $w-1$, $u-3$, $x-2$, $z-4$; but the graph has chromatic
number $3$.  Since $\overline{P}_6$ is a member of many families of
perfect graphs (such as brittle graphs, weakly chordal graphs,
perfectly orderable graphs, etc; see \cite{ramree01} for the
definitions), this algorithm will not perform optimally on these
classes.

%%%%%% CLIQUE
\section{Finding a maximum clique}

Given a coloring of a graph, there is a greedy algorithm that chooses
one vertex of each color in an attempt to find a clique of the same
size.

\begin{quote}
{\sc Algorithm Clique}

{\it Input:} A graph $G$ and a coloring of its vertices using $\ell$
colors.

{\it Output:} A set $Q$ that consists of $\ell$ vertices of $G$.

{\it Initialization:} Set $Q:=\emptyset$;

{\it General step:} For $c=\ell, \ldots, 1$ do:\\
Select a vertex $x$ of color $c$ that maximizes $N(x)\cap Q$, do $Q:=
Q\cup \{x\}$.
\end{quote}

Algorithm {\sc Clique} can be implemented in time ${\mathcal O}(m+n)$.

We claim that when the input consists of a Meyniel graph $G$ with the
coloring produced by {\sc LexColor}, then the output $Q$ of Algorithm
{\sc Clique} is a clique of size $\ell$.  This result is a consequence of
the next section, we show that when the output of the algorithm is not
a clique, we can find a Meyniel obstruction.

%%%%%
\section{Finding a Meyniel obstruction}
\label{obstruction}

Let $G$ be a general (not necessarily Meyniel) graph on which
Algorithm {\sc LexColor} is applied.  Let $\ell$ be the total number
of colors used by the algorithm.  Then we apply Algorithm {\sc
Clique}.  At each step, we check whether the selected vertex $x$ of
color $c$ is adjacent to all of $Q$ (this can be done without
increasing the complexity of Algorithm {\sc Clique} by maintaining a
counter which for each vertex counts the number of its neighbors in
$Q$).  If this holds at every step, then the final $Q$ is a clique
of cardinality $\ell$, and so we have a clique and a coloring of the
same size, which proves the optimality of both.  If not, then
Algorithm {\sc Clique} stops the first time $Q\cup \{x\}$ is not a
clique and records the current color $c$ and the current clique $Q$.
So we know that no vertex colored $c$ is adjacent to all of $Q$. Let
us show now how to find a Meyniel obstruction in $G$.  As usual, a
path is called odd or even if its length (number of edges) is
respectively odd or even.

% Starting from the fact that no vertex colored $c$ is adjacent to all
% of $Q$, we find an odd path $R$ in $G^*_i$ with a certain property (we
% call this a bad path), and a vertex $z$ adjacent to both ends of $R$.
% If it happens that $R$ has only vertices of $G$, then $R\cup\{z\}$
% induces what we call a near-obstruction, and in this case
% Lemma~\ref{lem:near-o} below shows how to obtain an obstruction in
% $G$.  If $R$ contains a contracted vertex, then Lemma~\ref{lem:badp}
% shows how to obtain either a near-obstruction in $G$ or a bad path in
% $G^*_j$ for some $j<i$, and so, by induction, we will eventually end
% up with an obstruction in $G$.  Now we describe this procedure
% formally.

% For each color $c\in\{1, \ldots, l\}$,
% let $k_c$ be the number of vertices colored $c$.  Then every vertex of
% $G$ can be renamed $x_i_c$, where $c\in \{1, \ldots, l\}$ is the color
% assigned to the vertex by the algorithm and $i\in \{1, \ldots, k_c\}$
% is the integer such that $x_i_c$ is the $i$-th vertex colored $c$.
% Thus $V(G)= \{x_1^1, x_1^2, \ldots, x_1^{k_1}, x_2^1,\ldots,
% x_2^{k_2},$ $ \ldots, x_l^1, \ldots, x_l^{k_l}\}$.  Let
% $G^*=G\setminus \{x_1^1, \ldots, x_1^{k_1}, \ldots, x_{c-1}^1, \ldots,
% x_{c-1}^{k_{c-1}}\}$, and for $1 \le i \le k_c$, let $G^*_i$ be the
% graph obtained from $G^*$ by removing $x_c^1,\ldots, x_c^i$ and adding
% a new vertex $w_c^i$ with edges to $N(x_c^1)\cup\cdots\cup N(x_c^i)$
% (in other words, vertices $x_c^1,\ldots, x_c^i$ are contracted to
% become $w_c^i$).

Let $n_c$ be the number of vertices colored $c$, and for $i=1, \ldots,
n_c$ let $x_i$ be the $i$-th vertex colored $c$.  Let $G^*$ be the
subgraph of $G$ obtained by removing the vertices of colors $<c$.  Let
$G^*_i$ be the graph obtained from $G^*$ by removing $x_1,\ldots, x_i$
and adding a new vertex $w_i$ with an edge to every vertex that is
adjacent to one of $x_1, \ldots, x_i$ (in other words, vertices $x_1,
\ldots, x_i$ are contracted into $w_i$).

Let $h\le n_c$ be the smallest integer such that every vertex of color
$>c$ has a neighbor in $\{x_1, \ldots, x_h\}$.  Integer $h$ exists
because $n_c$ has that property.  There is a vertex $a$ of $Q$ that is
not adjacent to $x_h$, because $x_h$ is not adjacent to all of $Q$.
Thus $h\ge 2$.  Note that $a$ is adjacent to $w_{h-1}$ in $G^*_{h-1}$.
There is a vertex $b$ of $Q$ that is adjacent to $x_h$ and not to
$w_{h-1}$, by the definition of $h$.  Then $w_{h-1}$-$a$-$b$-$x_h$ is
a chordless odd path in $G^*_{h-1}$.

For any $i>1$, a \emph{bad path} is any odd path $P=
w_{i-1}$-$v_1$-$\cdots$-$v_p$ in $G^*_{i-1}$ such that $v_p=x_i$, path
$P$ has at most one chord, and such a chord (if any) is
$v_{t-1}v_{t+1}$ with $1< t< p-1$.  Note that the path
$w_{h-1}$-$a$-$b$-$x_h$ obtained at the end of the preceding paragraph
is a bad path.

A \emph{near-obstruction} in $G$ is any pair $(P, z)$, where $P$ is a
path $v_0$-$\cdots$-$v_p$, with odd $p\ge 3$, $P$ has at most one
chord, such a chord (if any) is $v_{t-1}v_{t+1}$ with $0< t< p-1$,
vertex $z$ is a vertex of $G\setminus P$ that is adjacent to both
$v_0, v_p$, and the pair $(P, z)$ satisfies one of the following
conditions: \\
Type~1: $v_0v_2$ is the only chord of $P$, and $z$ is not adjacent to
either of $v_1, v_2$.  \\
Type~2: $v_1v_3$ is the only chord of $P$, and $z$ is not adjacent to
one of $v_1, v_3$.  \\
Type~3: $v_0v_2$ is not a chord of $P$, and $z$ is not adjacent to
$v_1$.  \\
Type~4: $v_0v_2$ and $v_1v_3$ are not chords of $P$, and $z$ is
adjacent to $v_1$ and not to $v_2$.  \\

The following lemmas show that the existence of a bad path is a
certificate that the graph is not Meyniel.  The proof of the first
lemma can easily be read as a linear-time algorithm which, given a bad
path, finds explicitly a near-obstruction.  Likewise, the proof of the
second lemma can easily be read as a linear-time algorithm which,
given a near-obstruction, finds explicitly an obstruction.  Since
$G^*_{h-1}$ contains the bad path $w_{h-1}$-$a$-$b$-$x_h$, these two
lemmas imply that $G$ contains a Meyniel obstruction.

%%%% LEMMA BAD PATH
\begin{lemma}
\label{lem:badp}
If $G^*_{i-1}$ contains a bad path, then $G$ contains a
near-obstruction.
\end{lemma}
%
%%% LEMMA near-obs
\begin{lemma}\label{lem:near-o}
If $G$ has a near-obstruction $(P, z)$, then $G$ has a Meyniel
obstruction contained in the subgraph induced by $P\cup\{z\}$.
\end{lemma}

\emph{Proof of Lemma~\ref{lem:badp}.} Let $P=
w_{i-1}$-$v_1$-$\cdots$-$v_p$ be a bad path in $G^*_{i-1}$, with the
same notation as above.  We prove the lemma by induction on $i$.  We
first claim that:
\begin{quote}
(*) There exists a vertex $z$, colored before $x_i$ with a color $>c$,
that is adjacent to $x_i$ and to $w_{i-1}$ in $G^*_{i-1}$ and
satisfies the following property.  If $v_1v_3$ is the chord of $P$,
then $z$ is not adjacent to at least one of $v_1$ and $v_3$.  If
$v_1v_3$ is not a chord of $P$, then $z$ is not adjacent to at least
one of $v_1$ and $v_2$.
\end{quote}
For let us consider the situation when the algorithm selects $x_i$ to
be colored.  Let $U$ be the set of vertices of $G^*_{i-1}$ that are
already colored at that moment.  We know that every vertex of
$G^*_{i-1}$ will have a color from $\{c, c+1, \ldots, \ell\}$ when the
algorithm terminates.  So, if $c\ge 2$, every vertex $v$ of $U$
satisfies $\forall\ c'<c$, $label_v(c') \neq 0$.  For any $X\subseteq
U$, let $color(X)$ be the set of colors of the vertices of $X$.   %
Put $T = N(x_i)\cap U$.  Every vertex of $T$ has a color $\geq c+1$,
and so is adjacent to at least one vertex colored $c$ in $G$ and
thus is adjacent to $w_{i-1}$ in $G^*_{i-1}$.  Specify one vertex
$v_r$ of $P$ as follows: put $r=3$ if $v_1 v_3$ is a chord of $P$;
else put $r=2$.  Note that $v_r$ is not adjacent to $w_{i-1}$ and
$v_r\neq x_i$ by the definition of bad path.  Suppose the claim is
false: so every vertex of $T$ is adjacent to $v_1$ and $v_r$.

Since every vertex of $T$ is adjacent to $v_1$, we have
$label_{v_1}(c') \geq label_{x_{i}}(c')$ for every color $c'>c$.
Since $v_1$ is adjacent to $w_{i-1}$, we have $label_{v_1}(c)> 0$.
Since $x_{i}$ is colored $c$, we have $label_{x_{i}}(c)=0$.  So
$label_{v_1}>_{Lex} label_{x_{i}}$, which means that $v_1$ is already
colored.  Moreover, $color(v_1)$ $\notin \{1, \ldots, c\}\cup
color(T)$.

Since every vertex of $T$ is adjacent to $v_r$, we have
$label_{v_r}(c')\geq label_{x_{i}}(c')$ for every color $c'>c$.  Since
$v_r$ is adjacent to $v_1$, we have $label_{v_r}(color(v_1))> 0$.
Since $color(v_1)$ $\notin \{1, \ldots, c\}\cup color(T)$ we have
$label_{x_{i}}(color(v_1))=0$.  So $label_{v_r}>_{Lex} label_{x_{i}}$,
which means that $v_r$ is already colored.  However, $v_r$ is not
adjacent to $w_{i-1}$, so $c$ was the smallest color available for
$v_r$ when it was colored, which contradicts the definition of
$w_{i-1}$ and $x_{i}$.  This completes the proof of Claim (*).

Now let $z$ be a vertex given by Claim (*).
% Vertex $z$ is adjacent to
% $w_{i-1}$ in $G^*_{i-1}$ because $z$ was colored before $x_i$ and
% received a color strictly greater than $c$.
(It takes time $deg(x_i)$ to find such a vertex $z$.)

Let $j$ be the smallest integer such that both $v_1$ and $z$ have a
neighbor in $\{x_1, \ldots, x_j\}$.  Then $j<i$ because $z$ and $v_1$
are adjacent to $w_{i-1}$.

Suppose that $x_j$ is adjacent to both $v_1$ and $z$.  Then
$(x_j$-$v_1$-$\cdots$-$v_p, z)$ is a near-obstruction in $G$.  Indeed,
by Claim (*), it is a near-obstruction of Type~2 if $v_1v_3$ is the
chord of $P$, of Type~3 if $v_1v_3$ is not a chord of $P$ and $z$ is
not adjacent to $v_1$, or of Type~4 if $v_1v_3$ is not a chord of $P$
and $z$ is adjacent to $v_1$ (and thus is not adjacent to $v_2$).

Now suppose that $x_j$ is not adjacent to both $v_1$ and $z$.  Then
the definition of $j$ implies that $j>1$ and either (a) $z$ is
adjacent to $x_j$ and not to $w_{j-1}$, and $v_1$ is adjacent to
$w_{j-1}$ and not to $x_j$ or (b) $v_1$ is adjacent to $x_j$ and not
to $w_{j-1}$, and $z$ is adjacent to $w_{j-1}$ and not to $x_j$.  In
either case, let $k$ be the smallest integer with $k\ge 1$ such that
$z$ is adjacent to $v_k$.  Such a $k$ exists because $z$ is adjacent
to $v_p$.
% (It takes time $deg(z)$ to compute $k$.)

Suppose that $k$ is odd.  If (a) holds, then let $P'=
w_{j-1}$-$v_1$-$\cdots$-$v_k$-$z$-$x_j$; if (b) holds, let
$P'=w_{j-1}$ -$z$-$v_k$-$\cdots$-$v_1$-$x_j$. Then $P'$ has at most
one chord, which is the chord of $P$ if it exists and if its two
end-vertices are in $P'$, so $P'$ is a bad path in $G^*_{j-1}$, and
the result follows by induction.

Now suppose that $k$ is even.  Then $k<p$ since $p$ is odd.  We
consider the following cases:

\emph{Case 1: $P$ has a chord $v_{t-1}v_{t+1}$ with $t<k$.} If (a)
holds, then let $P'= w_{j-1}$-$v_1$-$\cdots
$-$v_{t-1}$-$v_{t+1}$-$\cdots $-$v_k$-$z$-$x_j$; if (b) holds, let
$P'= w_{j-1}$-$z$-$v_k$-$\cdots $-$v_{t+1}$-$v_{t-1}$-$\cdots
$-$v_1$-$x_j$. Then $P'$ is chordless, so $P'$ is a bad path in
$G^*_{j-1}$, and the result follows by induction.

\emph{Case 2: $P$ has a chord $v_{k-1}v_{k+1}$.} When $z$ is
adjacent to both $v_{k+1}$ and $v_{k+2}$, if (a) holds, then let
$P'=
w_{j-1}$-$v_1$-$\cdots$-$v_{k-1}$-$v_{k+1}$-$v_{k+2}$-$z$-$x_j$; if
(b) holds, let $P'=
w_{j-1}$-$z$-$v_{k+2}$-$v_{k+1}$-$v_{k-1}$-$\cdots $-$v_1$-$x_j$; in
either case, $P'$ has only one chord, which is $zv_{k+1}$, so $P'$
is a bad path in $G^*_{j-1}$,  and the result follows by induction.
When $z$ is not adjacent to $v_{k+1}$, then $v_{k}$-$\cdots$-$v_p$
is a chordless path, so $(v_{k}$-$\cdots$-$v_p, z)$ is a
near-obstruction of Type~3. When $z$ is adjacent to $v_{k+1}$ and is
not adjacent to $v_{k+2}$, then $v_{k}$-$\cdots$-$v_p$ is a
chordless path, so $(v_{k}$-$\cdots$-$v_p, z)$ is a near-obstruction
of Type~4.

\emph{Case 3: $P$ has a chord $v_{k}v_{k+2}$.} When $z$ is adjacent
to $v_{k+1}$, if (a) holds, then let $P'=
w_{j-1}$-$v_1$-$\cdots$-$v_k$-$v_{k+1}$-$z$-$x_j$; if (b) holds, let
$P'= w_{j-1}$-$z$-$v_{k+1}$-$v_k$-$\cdots$-$v_1$-$x_j$; in either
case, $P'$ has only one chord, which is $zv_k$, so $P'$ is a bad
path in $G^*_{j-1}$, and the result follows by induction. When $z$
is not adjacent to $v_{k+1}$ and is adjacent to $v_{k+2}$, if (a)
holds, then let $P'=
w_{j-1}$-$v_1$-$\cdots$-$v_k$-$v_{k+2}$-$z$-$x_j$; if (b) holds, let
$P'= w_{j-1}$-$z$-$v_{k+2}$-$v_k$-$\cdots$-$v_1$-$x_j$; in either
case, $P'$ has only one chord, which is $zv_k$, so $P'$ is a bad
path in $G^*_{j-1}$,  and the result follows by induction. When $z$
is not adjacent to $v_{k+1}$ or to $v_{k+2}$, then
$(v_{k}$-$\cdots$-$v_p, z)$ is a near-obstruction of Type~1.

\emph{Case 4: $P$ has no chord $v_{t-1}v_{t+1}$ with $t\leq k+1$.}
When $z$ is adjacent to $v_{k+1}$, if (a) holds, then let $P'=
w_{j-1}$-$v_1$-$\cdots$-$v_k$-$v_{k+1}$-$z$-$x_j$; if (b) holds, let
$P'= w_{j-1}$-$z$-$v_{k+1}$-$v_k$-$\cdots$-$v_1$-$x_j$; in either
case, $P'$ has only one chord, which is $zv_k$, so $P'$ is a bad
path in $G^*_{j-1}$, and the result follows by induction. When $z$
is not adjacent to $v_{k+1}$, then $v_{k}$-$\cdots$-$v_p$ has at
most one chord, which is the chord of $P$ if it exists and if its
two end-vertices are in $P'$, so $(v_{k}$-$\cdots$-$v_p, z)$ is a
near-obstruction of Type~3.   This completes the proof of the last
case. \qed

\ 

Let us discuss the complexity of the algorithmic variant of this
proof.  When we find a new bad path, the value of $i$ decreases by
at least $1$, and so this happens at most $n_c$ times.  Dealing with
one bad path takes time ${\mathcal O}(deg(x_i)+deg(z))$ (for the
corresponding $i$), and $x_i$ is different at each call since $i$
decreases.  Vertex $z$ is also different at each call, because %
% if $z$ is not adjacent to $w_{j-1}$ then $z$ will not be adjacent to
% any future vertex $x_h$, and otherwise
$z$ becomes a vertex of the new bad path.  When the algorithm
produces a new bad path to be examined, it also tells if the path
has no chord or one chord, and what the chord is (if it exists); so
we do not have to spend any time to find this chord. So the total
complexity of this algorithm is $\mathcal{O}(m+n)$.

\

%%% PROOF OF LEMMA NEAR-OBSTRUCTION
\emph{Proof of Lemma~\ref{lem:near-o}.} We use the same notation for
$P$ as above.  We prove the lemma by induction on $p$.  If $p=3$, then
the hypothesis implies immediately that $P\cup\{z\}$ induces an
obstruction.  Now let $p\ge 5$.  If $(P, z)$ is a near-obstruction of
Type~1, 2 or 3, then let $r$ be the smallest integer $\ge 1$ such that
$z$ is adjacent to $v_r$.  If $(P, z)$ is of Type~4, then let $r$ be
the smallest integer $\ge 3$ such that $z$ is adjacent to $v_r$.

%% TYPE 1
First assume that $(P, z)$ is a near-obstruction of Type~1.  So $r\geq
3$.  If $r$ is odd, then $z, v_0, \ldots, v_r$ induce an odd cycle
with only one chord $v_0v_2$.  If $r$ is even, then $z, v_0, v_2,
\ldots, v_r$ induce an odd hole.

%% TYPE 2
Now assume that $(P, z)$ is a near-obstruction of Type~2.

\emph{Case 2.1: $z$ is not adjacent to either of $v_1, v_2$.} So
$r\geq 3$.  If $r$ is odd, then $z, v_0, \ldots, v_r$ induce an odd
cycle with only one chord $v_1v_3$.  If $r$ is even, then $r\geq 4$,
and $z, v_0, v_1, v_3, \ldots, v_r$ induce an odd hole.

\emph{Case 2.2: $z$ is not adjacent to $v_1$ and is adjacent to
$v_2$.} So $r=2$.  If $z$ is not adjacent to $v_3$, then $z, v_0,
v_1, v_3, v_2$ induce an odd cycle with only one chord $v_1v_2$.  So
suppose $z$ is adjacent to $v_3$.  If $z$ is also adjacent to $v_4$,
then $z, v_0, v_1, v_3, v_4$ induce an odd cycle with only one chord
$zv_3$.  If $z$ is not adjacent to $v_4$, then $p\ge 5$.  Consider
the path $P'=v_2$-$\cdots$-$v_p$. Then $P'$ is chordless and
$|P'|=|P|-r$, so  $(P', z)$ is a near-obstruction of Type~4, and the
result follows by induction.

\emph{Case 2.3: $z$ is adjacent to $v_1$.} So $r=1$, $z$ is not
adjacent to $v_3$ by the definition of Type~2, and so $p\ge 5$.
Consider the path $P'= v_1$-$v_3$-$\cdots$-$v_p$. Then $P'$ is
chordless and $|P'|=|P|-r-1$, so $(P', z)$ is a near-obstruction of
Type~3, and the result follows by induction.

%% TYPE 3
Now assume that $(P, z)$ is a near-obstruction of Type~3.  If $r$ is
odd, then $z, v_0, \ldots, v_r$ induce an an odd cycle with at most
one chord.  If $r$ is even, we consider the following cases:

\emph{Case 3.1: $P$ has a chord $v_{t-1}v_{t+1}$ with $t<r$.} Then $z,
v_0, \ldots, v_{t-1}, v_{t+1}, \ldots, v_r$ induce an odd hole.

\emph{Case 3.2: $P$ has a chord $v_{r-1}v_{r+1}$.} If $z$ is not
adjacent to $v_{r+1}$, then $z, v_0, \ldots, v_{r-1}, v_{r+1},
v_{r}$ in this order induce an odd cycle with only one chord
$v_{r-1}v_r$. So suppose $z$ is adjacent to $v_{r+1}$.  If $z$ is
also adjacent to $v_{r+2}$, then $z$, $v_0, \ldots, v_{r-1}$,
$v_{r+1}$, $v_{r+2}$ induce an odd cycle with only one chord
$zv_{r+1}$.  If $z$ is not adjacent to $v_{r+2}$, then $p\ge r+3$.
Consider the path $P'= v_{r}$-$\cdots$-$v_p$. Then $P'$ is chordless
and $|P'|=|P|-r$, so $(P', z)$ is a near-obstruction of Type~4, and
the result follows by induction.

\emph{Case 3.3: $P$ has a chord $v_{r}v_{r+2}$.} If $z$ is adjacent
to $v_{r+1}$, then $z, v_0, \ldots, $ $v_{r+1}$ induce an odd cycle
with only one chord $zv_r$.  So suppose $z$ is not adjacent to
$v_{r+1}$. If $z$ is adjacent to $v_{r+2}$, then $z, v_0, \ldots,
v_r, v_{r+2}$ induce an odd cycle with only one chord $zv_r$.  If
$z$ is not adjacent to $v_{r+2}$, then $p\ge r+3$.  Consider the
path $P'= v_{r}$-$\cdots$-$v_p$.  Then $v_rv_{r+2}$ is the unique
chord of $P'$ and $|P'|=|P|-r$, so $(P', z)$ is a near-obstruction
of Type~1, and the result follows by induction.

\emph{Case 3.4: $P$ has no chord $v_{t-1}v_{t+1}$ with $t\leq r+1$.}
If $z$ is adjacent to $v_{r+1}$, then $z, v_0, \ldots, v_{r+1}$
induce an odd cycle with only one chord $zv_r$.  If $z$ is not
adjacent to $v_{r+1}$, then consider the path $P'=
v_{r}$-$\cdots$-$v_p$. Then $P'$ has at most one chord, which is the
chord of $P$ (if it exists) and $|P'|=|P|-r$, so $(P', z)$ is a
near-obstruction of Type~3, and the result follows by induction.

% TYPE 4
Now assume that $(P, z)$ is a near-obstruction of Type~4.

Suppose that $P$ has a chord $v_{t-1}v_{t+1}$ with $2<t<r$.  If $r$ is
odd, then $z, v_1, \ldots, v_{t-1}, v_{t+1}, \ldots, v_r$ induce an
odd hole.  If $r$ is even, then $z, v_1, \ldots, v_r$ induce an odd
cycle with only one chord $v_{t-1}v_{t+1}$.

Now $P$ has no chord $v_{t-1}v_{t+1}$ with $2<t<r$.  If $r$ is odd,
then $z, v_0, \ldots, v_r$ induce an odd cycle with only one chord
$zv_1$.  If $r$ is even, then $z, v_1, \ldots, v_r$ induce an odd
hole.  This completes the proof of the four cases.  \qed

\

In the algorithmic variant of this proof, each recursive call
happens with the same vertex $z$, so we need only run once through
the adjacency array of $z$. Note that the first near-obstruction is
produced by the algorithm of Lemma~\ref{lem:badp}, so we already
know if $P$ has no chord or one chord, and what its chord is, if it
exists.
Computing the value of $r$ takes time ${\mathcal O}(r)$, and the
rest of each call takes constant time.  At each call, either a
Meyniel obstruction is output, or a near-obstruction $(P', z)$ is
obtained. Note that $|P'|\le|P|-r$, and we know if $P'$ has no chord
or one chord, and what its unique chord is (if it exits); so we do
not have to spend any time to find this chord.  So the total running
time is $\mathcal{O}(|P|+{deg}_P(z))$.

Algorithms {\sc LexColor} and {\sc Clique} run in time $\mathcal
O(n^2)$ and $\mathcal O(n+m)$ respectively, so the total time for
finding, in any graph, either a clique and coloring of the same size,
or a Meyniel obstruction is $\mathcal O(n^2)$.

\

\emph{Remark:} As mentioned earlier, on the graph $\overline{P}_6$
with vertices $u, v, w, x, y, z$ and non-edges $uv, vw, wx, xy, yz$, a
possible execution of {\sc LexBFS} colors the vertices in the
following order and with the given color: $v-1$, $y-2$, $w-1$, $u-3$,
$x-2$, $z-4$.  On this coloring, Algorithm {\sc Clique} will stop when
$c=1$ and $Q=\{x,u,z\}$.  No vertex of color $1$ is adjacent to all
$Q$ : $w$ is not adjacent to $x$ and $v$ is not adjacent to $u$.  So
$w$-$u$-$x$-$v$ is a chordless path of length $3$ between $w$ and $v$.
Vertex $w$ was colored before $x$ because of $y$, which is not
adjacent to $x$, and $\{w, u, x, v, y\}$ induces a Meyniel
obstruction.

%%%%%%
\section{Strong stable sets}
\label{stable}

It can be proved that, in the case of a Meyniel graph, the set of
vertices colored $1$ by Algorithm {\sc LexColor} is a strong stable
set.  But there are non-Meyniel graphs for which Algorithm {\sc
LexColor} and Algorithm {\sc Clique} give a coloring and a clique of
the same size but none of the color classes of the coloring is a
strong stable set (see the example at the end of this section).  In
that case we would like to be able to find a Meyniel obstruction.
This can be done in time ${\mathcal O} (n^3)$ as described below.

\begin{lemma}
\label{nice} Every nice stable set is a strong stable set.
\end{lemma}

{\it Proof.} Let $S=\{x_1,\ldots,x_k\}$ be a nice stable set of a
graph $G$.  Suppose there exists a maximal clique $Q$ with $Q\cap
S=\emptyset$.  Let $G^i$ be the graph obtained from $G$ by contracting
$x_1,\ldots, x_i$ into $w_i$.  For $i=1, \ldots, k$, consider the
following Property $P^i$: ``In the graph $G^i$, vertex $w_i$ is
adjacent to all of $Q$.''  Note that Property $P^k$ holds by the
maximality of $S$ and by the definition of $w_k$ and that Property
$P^1$ does not hold by the maximality of $Q$.  So there is an integer
$i\in\{2, \ldots, k\}$ such that $P^i$ holds and $P^{i-1}$ does not.
Vertex $x_i$ is not adjacent to all of $Q$ by the maximality of $Q$.
So, in the graph $G^{i-1}$, the clique $Q$ contains vertices $a$ and
$b$ such that $a$ is adjacent to $w_{i-1}$ and not to $x_i$ and $b$ is
adjacent to $x_i$ and not to $w_{i-1}$, and then the path
$w_{i-1}$-$a$-$b$-$x_i$ is a $P_4$, which contradicts the property
that $S$ is nice.  \qed

\

Now, for any graph $G$ and any vertex $v$ of $G$, we can find a
Meyniel obstruction or a strong stable set containing $v$ by the
following algorithm:

Apply the algorithm {\sc LexColor} on a graph $G$, choosing $v$ to be
the first vertex to be colored.  Let $S=\{s_1, \ldots, s_{n_1}\}$ be
the set of vertices colored $1$.  So $S$ is a maximal stable set.  We
can check in time ${\mathcal O}(n^3)$ whether $S$ is a nice stable
set.  If $S$ is a nice stable set, then $S$ is a strong stable set by
Lemma~\ref{nice}.  If $S$ is not a nice stable set, then the checking
procedure returns some $i\in \{2, \ldots, n_1\}$ such that there is an
induced path $t_{i-1}$-$a$-$b$-$s_i$, where $t_{i-1}$ is the vertex
obtained by contracting $s_1, \ldots, s_{i-1}$.  Applying the
procedure described in Lemmas~\ref{lem:badp} and~\ref{lem:near-o} of
Section \ref{obstruction} to this bad path $t_{i-1}$-$a$-$b$-$s_i$
gives a Meyniel obstruction in $G$.

\

\emph{Remark:} Here is an example of a non-Meyniel graph for which
Algorithm {\sc LexColor} followed by Algorithm {\sc Clique} can give a
coloring and a clique of the same size but none of the color classes
of the coloring is a strong stable set.  Consider the graph $G$ form
by the 3 triangles $\{a,d,e\}$, $\{b,f,g\}$, $\{c,h,i\}$ plus the
edges $af$, $ah$, $bd$, $bi$, $ce$, $cg$.  Algorithm {\sc LexColor}
can color the vertices in the following order and with the given
color: $d-1$, $b-2$, $f-1$, $g-3$, $c-1$, $i-3$, $h-2$, $a-3$, $e-2$,
and Algorithm {\sc Clique} returns the clique $\{a,d,e\}$.  The
algorithms give a coloring and a clique of the same size but none of
the color classes $\{c,d,f\}$, $\{b,e,h\}$ or $\{a,g,i\}$ is a strong
stable set.

\section{Comments}

The algorithms presented here are not recognition algorithms for
Meyniel graphs.  It can happen that the input graph is not Meyniel
and yet the output is a clique and a coloring of the same size.

The fastest known recognition algorithm for Meyniel graph is due
to Roussel and Rusu \cite{rourus-rec} and its complexity is
${\mathcal O}(m(m+n))$, (where $n$ is the number of vertices and
$m$ is the number of edges), which beats the complexity of the
algorithm of Burlet and Fonlupt \cite{burfon}. So it appears to be
easier to solve the Meyniel Graph Robust Algorithm Problem than to
recognize Meyniel graphs. It could be the same for perfect graphs:
it might be simpler to solve the Perfect Graph Robust Algorithm
Problem than to recognize perfect graphs. Currently, the
recognition of perfect graphs is done by an ${\mathcal O}(n^9)$
algorithm due to Chudnovsky, Cornu\'ejols, Liu, Seymour and Vu\v
skovi\'c \cite{chu} which actually recognizes Berge graphs (graphs
that do not contain an odd hole or an odd antihole). The class of
Berge graphs is exactly the class of perfect graphs by the Strong
Perfect Graph Theorem of Chudnovsky, Robertson, Seymour and Thomas
\cite{CRST}.

%\clearpage

\end{document}